\documentclass[sigconf]{acmconference}

\usepackage{amsmath}
\usepackage{fancyvrb}
\usepackage{xspace}
\usepackage{booktabs}
\usepackage[capitalise]{cleveref}

\usepackage{enumitem}
\usepackage[all]{nowidow}
\usepackage{siunitx}
\usepackage{balance}

\fvset{fontsize=\footnotesize}

\setlist[description]{labelsep=1ex, labelwidth=0em, leftmargin=3ex, topsep=0ex, partopsep=1ex, itemsep=0ex, parsep=0ex}

\usepackage{tikz}
\usetikzlibrary{calc}

\newcommand{\benchmark}[1]{\textit{#1}}

\newcommand{\xsonel}{\mbox{XS1-L}\xspace}

\begin{document}

\setcopyright{acmcopyright}

\title{Data Dependent Energy Modeling for\\ Worst Case Energy Consumption
Analysis}
\author{James Pallister, Steve Kerrison, Jeremy Morse, Kerstin Eder\\
Department of Computer Science, University of Bristol, BS8 1UB, UK\\
firstname.lastname@bristol.ac.uk
}

\maketitle

\begin{abstract}

Safely meeting Worst Case Energy Consumption (WCEC) criteria requires
accurate energy modeling of software. We investigate the impact of
instruction operand values upon energy consumption in cacheless embedded
processors. Existing instruction-level energy models typically use
measurements from random input data, providing estimates unsuitable for safe
WCEC analysis.

We examine probabilistic energy distributions of instructions and propose a model for
composing instruction sequences using distributions, enabling WCEC analysis on
program basic blocks. The worst case is predicted with statistical
analysis. Further, we verify that the energy of embedded benchmarks can be characterised
as a distribution, and compare our proposed technique with other methods of
estimating energy consumption.

\end{abstract}

\section{Introduction}

In real-time embedded systems, execution time of a program must be bounded.
This can provide guarantees that tasks will meet hard deadlines and the system
will function without failure. Recently, efforts have been made to give upper
bounds on program energy consumption to determine if a task will complete
within an available energy budget \cite{Jayaseelan2006, isa-energy-lopstr13, scopes15, Wagemann2015, taco:GeorgiouKCE2017}. 
Motivating this research is the
developing Internet of Things (IoT) market, where pervasive deeply embedded
devices require battery- or harvester-based energy sources. Inaccurate
energy consumption estimates can lead to expensive maintenance, or task failure
if power thresholds are exceeded.

However, existing WCEC analyses often use energy
models that do not explicitly consider the dynamic power drawn by switching of
data in instruction operands, instead producing an upper bound using averaged
random or scaled
instruction models~\cite{Jayaseelan2006, isa-energy-lopstr13, scopes15, Wagemann2015, taco:GeorgiouKCE2017}. 
A safe and tightly bound model for WCEC analysis must be close to the
hardware's actual behaviour, but also give guarantees that it never
under-estimates. Current models have not been analysed in this context to
provide sufficient confidence, and power figures from manufacturer datasheets
are not sufficiently detailed to provide tight bounds.

Energy modeling allows the energy consumption of software
to be estimated without taking physical measurements. Models may assign
an energy value to each instruction~\cite{Tiwari1994,Kerrison2015}, to a predefined set of
processor modes~\cite{NunezYanez2013}, or use a detailed approach that
considers wider processor state, such as the data for each
instruction~\cite{Steinke2001}. Although measurements are typically more
accurate, models require no hardware instrumentation, are more versatile and
can be used in many situations, such as statically predicting energy
consumption~\cite{Jayaseelan2006,isa-energy-lopstr13,fopara15,scopes15,Wagemann2015,taco:GeorgiouKCE2017},
which significantly reduces the costs
and barriers to entry of energy estimation.

\begin{figure}
    \centering
    \includegraphics[width=1.0\linewidth,clip,trim=1.25cm 0.5cm 0.75cm 0.5cm]{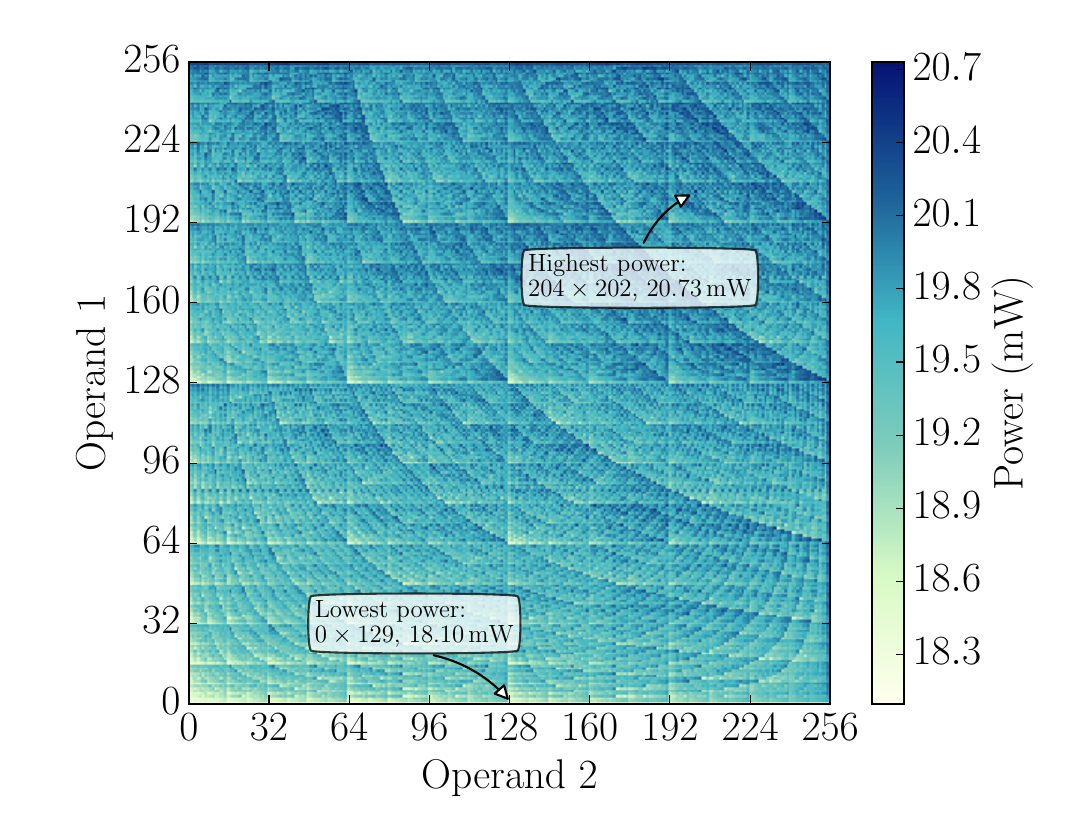}
    \caption{Power map of \texttt{mul} instruction for the AVR processor; total range is 15\,\% of SoC power.}
    \label{fig:avr_mul_heatmap}
\end{figure}

Changes in energy consumption caused by different data can have a significant
impact on the overall energy consumption of a program. Previous work on a
32-bit processor~\cite{Kerrison2015} used in this paper has reported up to
20\,\% difference in energy consumption due to data values, while other
processors may have up to 50\,\% of energy caused by data
switching~\cite{Ascia2001}.  In our own experiments we find 15\,\% difference
in a simple 8-bit AVR processor. This device has no caches, no OS and no high
power peripherals. This difference can be seen in \Cref{fig:avr_mul_heatmap},
which shows the power for a single cycle, 8-bit multiply instruction in this
processor.\footnote{All measurements in this paper are taken on physical hardware.} The diagram was constructed by taking hardware measurements for
every possible combination of eight bit inputs.  In this paper we choose to
focus on the contribution of operand data to WCEC as it has not been studied in
detail, whereas program length and execution path analysis is the subject of
much prior work. We wish to create energy models that can account for all
sources of energy, including data dependent sources.

Accounting for data dependent effects in an energy model is a challenging task,
which we split into two parts.
Firstly, the energy effect of an instruction's manipulation of processor state
needs to be modeled. This is an infeasible amount of data to exhaustively
collect. A 32-bit three-operand instruction has $2^{96}$ possible data value
combinations.

Secondly, a technique is required to derive the energy consumption for a
sequence of instructions from such a model.
The composition of data dependent instruction energy models is a particularly difficult task. The data causing maximum energy consumption for one instruction may minimise the cost in a subsequent, dependent instruction. Finding the greatest cost for such sequences requires searching for inputs that maximise a property after an arbitrary computation, which is again an infeasibly large task \cite{DBLP:journals/corr/MorseKE16}.
Over-approximating by summing the worst possible data dependent energy consumption of each instruction in a sequence, regardless of whether such a computation can occur, would lead to a significant overestimation of the upper bound.

We analyse individual instructions and explore probabilistic modeling approaches
to determine the maximum energy consumption of instruction sequences.
This provides a means to analyse complex programs at the block level.  The
analysis exposes how data correlations reduce maximum energy. A degenerate case
is discovered, where the sequence of instructions results in a bimodal energy
distribution.

We then explore the effect of data on the maximal energy consumption of programs,
performing probabilistic analysis on the distributions of energy consumption
obtained. Data's effect on entire programs is explored through several methods,
finding that random data forms distributions from which a maximum energy can be
estimated. To the best of our knowledge we believe that this is the first work
to statically estimate energy variability due to operand data in programs, or
to use probabilistic techniques to model data-dependent energy consumption.

This paper is organised as follows. The next section discusses related work.
Section~\ref{s:modelling} models individual and sequences of instructions in
the AVR, and \Cref{s:datadep} describes a sequence that causes a bimodal energy
distribution.  In Section~\ref{s:programs} the effect of data on two full
programs is explored for two processors. \Cref{s:discuss} discusses the
implications of our work, while Section~\ref{s:conclusion} concludes and gives
an outlook on future work.

\section{Related work}

Worst Case Execution Time (WCET) analysis attempts to find an upper bound on
the time taken for an arbitrary program to execute~\cite{Hahn2015,wcetsurvey}.
A key approach is a method called Implicit Path Enumeration Technique
(IPET)~\cite{Li1995a}, which estimates an upper bound given information about a
program's control flow graph. Of recent interest has been work on Worst Case
Energy Consumption (WCEC), utilising methods from WCET, combining them with
energy modeling to bound program energy consumption~\cite{Jayaseelan2006,
Wagemann2015,taco:GeorgiouKCE2017}. In many of these studies, energy models are not tailored to the
worst case, nor is the impact of data on energy consumption adequately
reflected. This can lead to unsafe results if the analysis is to be relied on
for guarantees of system behaviour within a given energy budget, as discussed extensively in~\cite{taco:GeorgiouKCE2017}, or bounds will be overly pessimistic if conservative
over-approximations are used to ensure safety, as 
identified in~\cite{Wagemann2015}.

A common form of model for embedded systems is an instruction level model. For
example, Tiwari et al.~\cite{Tiwari1996} use an energy cost for each
instruction and an energy cost for the circuit switching effect between each
instruction, as well as an extra term to cover other external system effects.

Steinke et al.~\cite{Steinke2001} construct a more detailed energy model that
does consider the effects of data as well as the instructions. The Hamming
distance between consecutive data values and their Hamming weights are
considered with respect to both register and memory access. The technique
achieves a 1.7\,\% average error, however every input must be known and
every internal state calculated under the technique, an infeasibly large
problem for reasoning about programs as a whole.
Park et al.~\cite{Park2011} consider how
different operand values affect the energy consumption, using a range of values
between \texttt{0x0000} and \texttt{0xFFFF} to ensure that there is a large
number of different Hamming distances between operands. Further studies have
extensively used the Hamming weight to account for data
energy~\cite{Sarta1999}. The study notes that the Hamming distance and weight
are particularly useful for subsequent values on busses in the processor, and
less useful for combinatorial instructions, such as arithmetic.

Ascia et al.~\cite{Ascia2001} build upon the approach, exploring how the data transitions from 0 to 1 and 1 to 0 can be given different energy costs. In a study of the Leon3 processor~\cite{Penolazzi2009}, taking data into account was found to reduce the model error when a `typical' number of switching bits was factored in.

Kojima et al.~\cite{Kojima1995} measure the data's effect on power of hardware
units such as
adder, multiplier and registers in a DSP. Register file power was found to show
linear dependence on the Hamming weight of operands, while the adder shows
moderate correlation with the Hamming distance between successive operands.
However, the multiplier shows very little correlation with Hamming distance,
except when one of the inputs is held constant. This supports the suggestion
that combinatorial blocks require parameters other than the Hamming distance
and weight~\cite{Sarta1999}. Similar conclusions have been reached in studies
which attempt to find the maximum power a circuit may trigger~\cite{Najm1994}.
Many studies attempt to maximise the power consumption of a circuit, using a
weighted maximum satisfiability approach~\cite{Devadas1992} and genetic
algorithms~\cite{Hsiao1999}.

The reachability of a particular state has large implications for maximum
energy consumption. Hsiao et al.~\cite{Hsiao1997} use a genetic algorithm (GA)
to
determine the maximum power per cycle in VLSI circuits. They show that the peak
power for a single cycle is higher than the peak power over a series of cycles,
as sequences of operations constrain the state transitions, making repeated
worst cases unreachable.  Therefore, for instructions, the data triggering
highest energy consumption in one may preclude subsequent instructions from
consuming their maximal energy.

GA techniques have also been used to estimate the WCEC
of certain software components: W\"agemann et al.~\cite{Wagemann2015} use GAs
to search for the maximum amount of energy a single instruction can consume.
Liqat et al.~\cite{fopara15} apply GAs to program basic blocks, to search for
the worst-case input of a sequence of instructions. Both require on-line
testing of the hardware, and yield an energy value that is likely to be close
to the WCEC for the instruction or block, however this cannot be guaranteed.

Probability theory has also been used to characterise how circuits dissipate
power. Burch et al.~\cite{Burch1993} take a Monte Carlo approach, simulating
the power of different input patterns to a circuit. The paper hypothesises that
the distribution of powers can frequently be approximated by a normal
distribution, as a consequence of the central limit theorem~\cite{Feller1945}.
While the central portions of the probability distribution fit well to a normal
distribution, the tails diverge, implying that a different distribution would
be a better fit when maximum power is of interest.
Studies have used extreme value theory to rectify this issue.
The extreme value
distribution is of importance when the maximum of a series of random variables
is needed. It has been used in maximum power estimation of VLSI
circuits~\cite{Evmorfopoulos2002}, giving a probabilistic estimate of maximum
power with a small number of simulations.
Further, the WCET community have used extreme value techniques to
probabilistically bound program execution times \cite{pwcet}.

In summary, energy consumption and data dependency have been considered at the
VLSI level using a variety of techniques. However, there has been little
exploration of data dependency at the instruction or application level for
worst case energy consumption analysis.
The study of data dependency is fundamental in establishing both
tighter WCEC upper bounds as well as providing greater confidence in the safety
of those bounds.
To support program level analysis, our challenge is to associate energy
consumption costs with software constructs, rather than hardware blocks.

\section{Instruction Modeling}\label{s:modelling}
\subsection{Individual instructions}

\begin{figure}[t]
    \includegraphics[width=\linewidth,clip,trim=0cm 0.4cm 0cm 0.4cm]{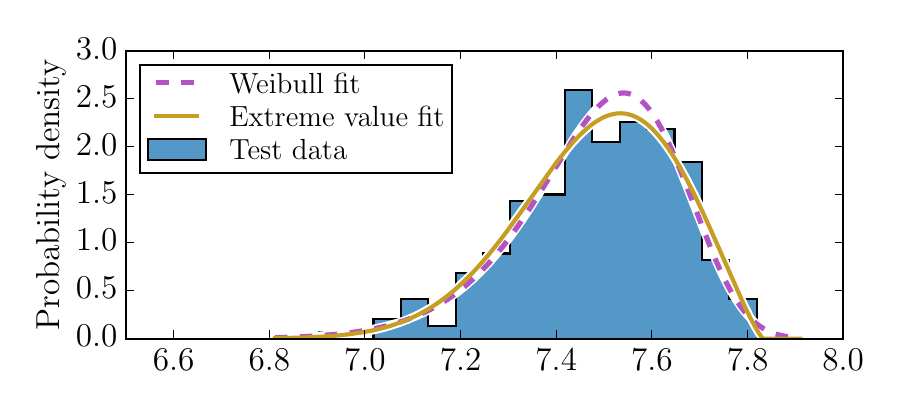}
    \caption{Distribution of energy (mJ) for \texttt{lsl}.}
    \label{fig:single_lsl}
\end{figure}

While characterising energy consumption of the entire state space of an
instruction is infeasible, requiring all combinations of all inputs to be
tested, we can statistically characterise an instruction's energy consumption
for a representative input sample. We use the AVR platform for our initial
experiments due to its simplicity, repeatedly running code sequences that load
random values (uniformly distributed) into registers and then executing an
instruction using those registers. An example of the resulting distribution of
energy values can be seen in \Cref{fig:single_lsl}.

We fit the Weibull distribution to the random data energy distribution, under
the hypothesis that the switching and hence power dissipation caused by random
data will be close to the maximum.
The distribution can then be examined to estimate an upper bound.
The Weibull cumulative distribution function is a stretched exponential,
allowing it to characterise many distributions with a high density followed
by a long tail.
This matches our hypothesis of most random-data power dissipation being close
to the maximum, followed by a tail of less-than-maximum samples.
\Cref{fig:single_lsl} shows the regular Weibull distribution and an extreme
value fit, the reversed Weibull distribution.
The latter fits best, however it operates on the premise that there is a
finite cut-off point (furthest right) beyond which there are no samples, which
we cannot guarantee.
This could lead to underestimation of an upper bound, compromising safety, and so we proceed with
the regular Weibull distribution.
The Weibull cumulative probability distribution  (CDF) is given by:
\begin{equation}
F(x; k, \mu, \sigma) = 1 - e^{-\left(\frac{x-\mu}{\sigma}\right)^k}.
\end{equation}
where
\begin{itemize}
	\item $x$ is the random variable,
	\item $k$ is the shape parameter, defining the extent to which the distribution is stretched,
	\item $\mu$ is the location parameter, defining the shift of the distribution,
	\item $\sigma$ is the scale parameter, defining the size of the distribution.
\end{itemize}

Using the distribution calculated, and the total size of the input data
space,\footnote{i.e., the finite set of all inputs an instruction may operate
upon} an estimate of the maximum possible average power can be calculated,
\begin{equation} (1-\text{CDF}(x))\cdot S = 1, \end{equation}
where $\text{CDF}(x)$ is the cumulative density function of the probability
distribution, and $S$ is the total size of the data space. Intuitively, this is
equivalent to finding the value of the percentile representing the highest
power dataset in the entire data space.

\subsection{Composing instructions}

To statistically characterise programs, our model must
support modeling sequences of instructions. The natural unit of such sequences
is a
\textit{basic block}, a sequence with a unique entry point and ending with a
branch instruction.
A simplistic method to generate basic block energy distribution would be to
measure each block in isolation with random input data. However, even for small
programs this is significant work that increases prohibitively with
program size.

A more tractable approach is to take the energy model for each instruction in a
basic block and compose these models. This
only requires each individual instruction to be characterised, after which any
size basic block can have its model deduced.

\begin{figure}[t]
    \includegraphics[width=\linewidth]{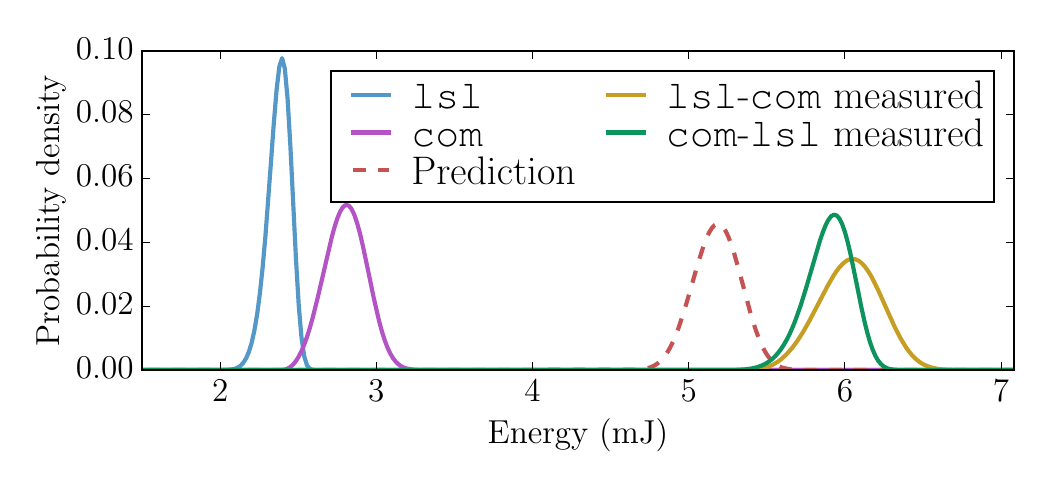}
    \caption{Comparison of predicted energy for the combination of a \texttt{com} and a \texttt{lsl} instruction with a simplistic model.}
    \label{fig:single_combinations}
\end{figure}

By convolving the individual distributions of instructions
together, a prediction of multiple instructions can be constructed.
Figure~\ref{fig:single_combinations} shows the distributions for the
instructions \texttt{com} (bitwise complement) and \texttt{lsl} (logical shift
left). The dashed curve shows the expected distribution. The two curves marked
in \textcolor[rgb]{0.0509803921569,0.476470588235,0.264705882353}{green} and
\textcolor[rgb]{0.606470588235,0.459607843137,0.117254901961}{orange} show the
actual distributions of the energy for each instruction --- one for
\texttt{com}, then \texttt{lsl}, and the second for \texttt{lsl}, then
\texttt{com}. These distributions are not similar, and more importantly are
higher than the prediction, resulting in an underestimate of the worst case
energy consumption.

The difference in distributions stems from the surrounding instructions --- to
evaluate the instructions, the sequence is prefixed with a \texttt{mov}
instruction to set up the values going into \texttt{com} and \texttt{lsl}. This
suggests that the actual switching of data between the instructions can have a
significant impact on not only the average energy, but the shape of the
distribution too.

\subsection{Instruction transitions}

In light of the inaccuracy of single instruction models,
we use a model based on Tiwari et al.~\cite{Tiwari1994}, using transition
distributions to represent the data dependent transition between instructions.
We assume that the energy consumption can be characterised using only
transitions and no instruction base costs. To calculate the program energy
$E_p$, formally,
\begin{align}
E_p &= \sum_{(i,j) \in TRACE_p} E_{i,j}\text{, where } E_{i,j}\sim\textrm{Weibull distribution}.
\end{align}

With $TRACE_p$ the sequence of instruction pairs $ISA \times ISA$
that make up the execution of the program $p$, ISA the set of all instructions,
and $E_{i,j}$ being the energy distribution of switching from instruction $i$
to $j$. $E_p$ can be calculated by convolving the individual probability
distributions,
\begin{align}
f_p(x) &=  \bigotimes_{(i,j) \in TRACE_p} f(x; k_{i,j}, \mu_{i,j}, \sigma_{i,j}).
\end{align}
where $\mu$, $\sigma$ and $k$ are the parameters for the Weibull probability
density function of each transition, $f$. The symbol $\bigotimes$ is the
convolution operator, and $f_p$ is the probability density function of the
instruction sequence. The convolution of two Weibull probability density
functions is not known to have an analytical solution, so it is solved
numerically for the purposes of this study: we project the distributions onto
histograms and then directly convolve the distributions, at the expense of some
sampling error.

\subsection{Data collection}

The collection of transition distributions for each pair of instructions is particularly challenging. The most simplistic approach is to repeat a pair of instructions with specified data and measure the energy, e.g.\
\begin{center}
  \begin{minipage}{0.6\linewidth}
  \begin{Verbatim}
  add r0, r1, r2;  sub r3, r4, r5.
  \end{Verbatim}
  \end{minipage}
\end{center}

However, after the first repetition \texttt{r0} and \texttt{r3} will not exhibit the same switching as they did in the first iteration --- the value in the register will not change. Therefore, register values should be randomised before and after the execution of the instructions,

\vspace{2ex}
\begin{center}
\begin{minipage}{0.6\linewidth}
\newcommand{\node}[1]{%
\begin{tikzpicture}[remember picture]%
\begin{scope}[]%
\coordinate (n#1);%
\end{scope}%
\end{tikzpicture}}
\begin{Verbatim}[numbers=left,numbersep=2.5em,commandchars=\|\§\! ]
mov r0, X        |node1
mov r3, Y        |node2
add r0, r1, r2   |node3
sub r3, r4, r5   |node4
...              |node5
\end{Verbatim}
\begin{tikzpicture}[overlay, remember picture]
    \newcommand{\drawb}[3]{
    \draw ($(#1)+(-0.2cm,-0.15ex)$) -- ($(#1)+(0,-0.15ex)$) -- ($(#2)+(0,+0.15ex)$) node[midway,anchor=west] {#3} -- +(-0.2cm,0);
    }
    \drawb{n1}{n2}{$E_{mov, mov}$}
    \drawb{n2}{n3}{$E_{mov, add}$}
    \drawb{n3}{n4}{$E_{add, sub}$}
    \drawb{n4}{n5}{$E_{sub, \,\ldots}$}
\end{tikzpicture}
\end{minipage}
\end{center}

\noindent where $X$ and $Y$ are unique registers containing independent,
uniform random variables. In addition to $X$ and $Y$, registers \texttt{r1},
\texttt{r2}, \texttt{r4} and \texttt{r5} are initialised to random variables.
This ensures all variables that could affect the transition distribution
between two instructions are random and should lead to each transition
distribution conforming to the Weibull distribution. The above test forms the
$E_{x,y}$ distributions seen to the right of the instructions. This approach
introduces additional \texttt{mov} instructions to the test. These are
convolved with the distribution that is of interest, so they must first be
found in order to then be eliminated.

A large number of values can then be assigned to all variables in the sequence,
$s$, and the energy, $E_s$ measured for each. For example, this can form the following
equation that must be solved to find $E_{add,sub}$,
\begin{equation}\label{e:e_s}
    E_s = E_{mov,mov}\otimes E_{mov,add} \otimes E_{add,sub} \otimes E_{sub,mov}.
\end{equation}

To solve, we first find the distribution for $E_{mov,mov}$, then the
distributions for $E_{mov,i}$, where $i$ is another instruction. For simplicity
it is assumed that $E_{i,j} \equiv E_{j,i}$. The $E_{mov,mov}$ distribution can
be found by finding the distribution for a repeating sequence of four
\texttt{mov}s, alternating between two destination registers and each using a
unique source register containing an independent uniform random variable. The
resulting distribution is $E_{mov,mov}$ convolved with itself four times.
This, along with similarly formed tests to find $E_{mov,i}$ and $E_{mov,j}$
yields the transition distribution for any $E_{i,j}$.

Currently, we have used this approach to obtain transition distributions for a
subset of the AVR's instruction set. For each instruction pairing we used
256 random number assignments for single operand instructions and 1024 for
two operand instructions.
For each assignment of random numbers, we ran an unrolled
loop of the code sequence 65536 times and took the average energy consumption
across this period. Each such measurement takes on average one second, and each
instruction pairing takes between five and twenty minutes to be characterised.

\subsection{Instruction sequence tests}

\begin{figure*}
	\centerline{
    \includegraphics[width=0.625\linewidth]{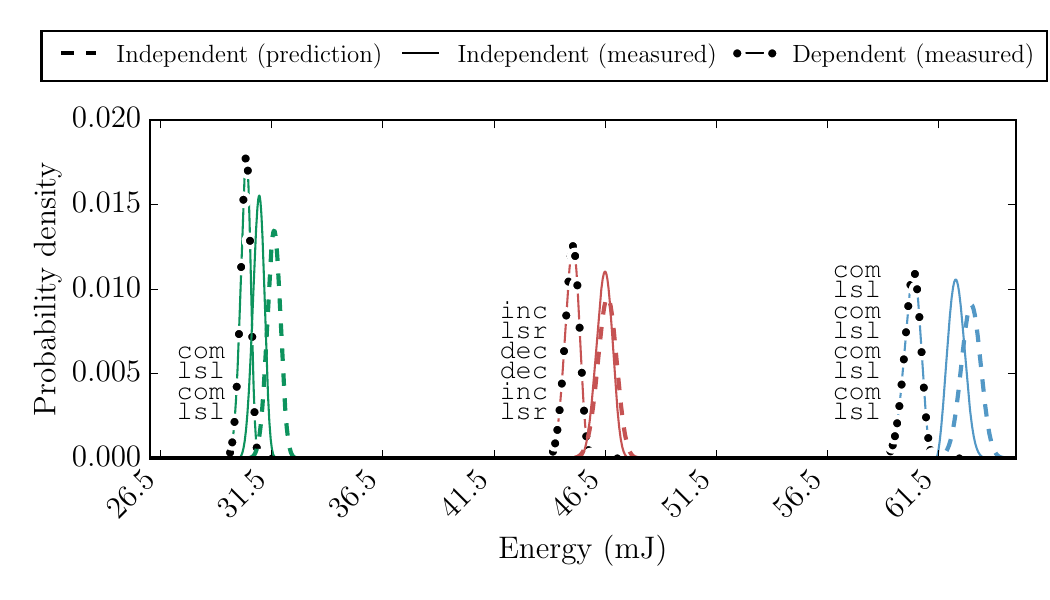}
	}
    \caption{Energy distributions of three sequences.}
    \label{fig:instruction_histograms}
\end{figure*}

Using the previously described method, \Cref{fig:instruction_histograms} shows
the predicted distributions for three short instruction sequences as dashed
lines. In all cases the prediction is conservative, with the mean of the
distribution overestimated. This makes it useful in a worst case energy model,
since the $99^\mathrm{th}$ percentile can be taken as a probabilistic
estimation of maximum energy, for example. While we test only arithmetic
instructions, we expect that loads, stores and branches can also be
characterised in the same way.

The figure also demonstrates the case where values in the registers are not
randomly distributed (dashed lines with points). Instead, they are dependent on
the results of previous instructions. All of these distributions have a smaller
mean, where the correlation between registers causes lower overall energy and
so the upper bound holds, as one would expect.

The tests in this section only showed arithmetic instructions. However, the
distributions for load and store instructions are similar and can be composed
similarly. It is expected that branch instructions will be simple to
characterise --- while there are often no direct inputs to a conditional
branch, the state of the control flags influences the direction of the branch.
Characterising the whole instruction set of a machine would mean testing each
pairing of instructions. This would require a significant number of experiments for large
machines, but is potentially feasible.

\section{Data dependency}
\label{s:datadep}

\begin{figure}
    \hspace*{-2mm}\includegraphics[width=1.05\linewidth, clip, trim=0cm 0.4cm 0cm 0.5cm]{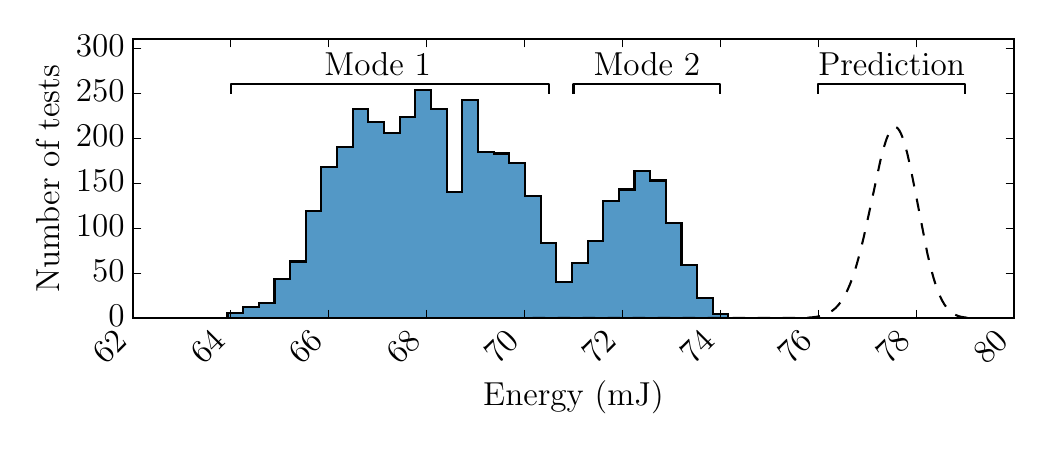}
    \caption{Histogram showing energy distribution for a sequence of multiplies.}
    \label{fig:bimodal_avr_histogram}
\end{figure}

The previous section suggests that effects of computation may impact the
location of the distribution. This section presents a case where this occurs in
certain sequences of multiplication. Consider a sequence of
\texttt{mul} and \texttt{mov} instructions calculating $a^{13}\cdot b^8$, where
$\texttt{r20}=a$ and $\texttt{r21} = b$:
\vspace{2ex}

\begin{center}
\begin{minipage}{0.5\linewidth}
    \newcommand{\firstnode}{\tikz[remember picture] \coordinate (n1);}
    \newcommand{\secondnode}{\tikz[remember picture] \coordinate (n2);}
    \begin{Verbatim}[numbers=left,numbersep=2.5em,commandchars=\|\§\! ]
mov r3, r20;  mov r4, r21;
mul r3, r4;   mov r2, r0;   |firstnode
mul r2, r3;   mov r4, r0;
mul r4, r2;   mov r3, r0;   |secondnode
    \end{Verbatim}
    \begin{tikzpicture}[remember picture, overlay]
        \draw ($(n1) + (-0.25cm,0.9ex)$) -- ($(n1) + (0,0.9ex)$) -- (n2) node [midway,rotate=0,right] {\small Repeat 2 times} -- +(-0.25cm,0);
    \end{tikzpicture}
\end{minipage}
\end{center}

Note that on AVR, \texttt{mul} implicitly writes to \texttt{r0}.

The sequence was measured for its energy under different inputs to produce the
histogram in \Cref{fig:bimodal_avr_histogram}. The distribution has two large
peaks, labelled with two modes. In this particular example, the lower energy
peak is caused by the computation collapsing to a zero value that persists
throughout the sequence. The upper mode is caused by neither of the inputs to
any of the multiplies being zero, which occurs for every multiply when both
inputs are odd.

While this type of behaviour will affect the tightness of the energy's
upper bound, it does not affect its safety, since it is the upper mode that is
captured by the prediction. Additionally we believe this circumstance, where
integer values overflow register sizes, is rare in real-world code as most
programming languages (particularly C) treat integer overflow as undefined
(i.e. erroneous) behaviour.

\section{Analysing whole programs}\label{s:programs}

\begin{figure*}
    \hspace*{-0.5cm}\includegraphics[width=1.04\linewidth]{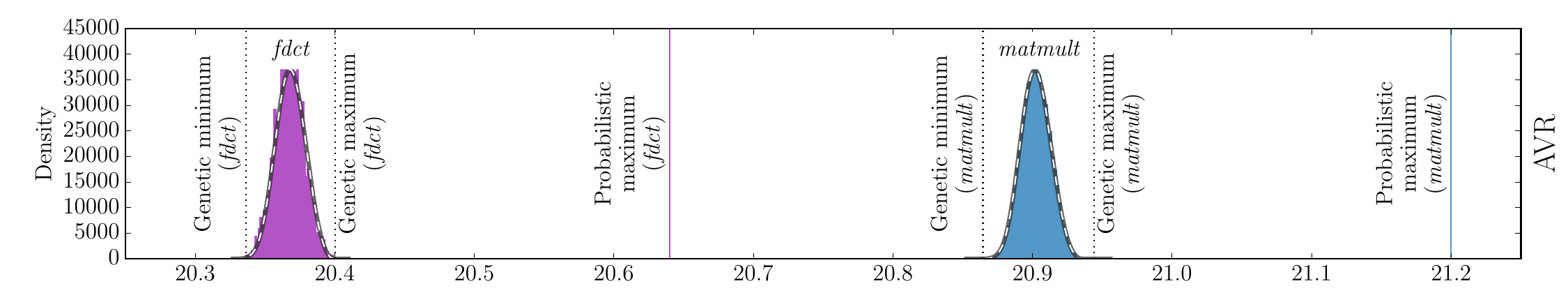}
    \hspace*{-0.5cm}\includegraphics[width=1.04\linewidth,clip,trim=0cm 0.5cm 0cm 0.5cm]{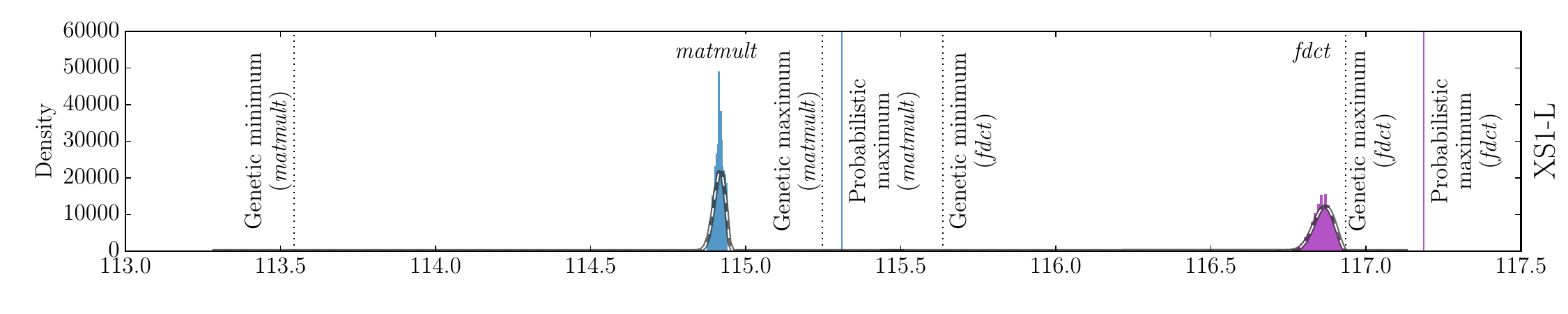}
    \caption{Power distributions (mW) for random datasets over full program benchmarks.}
    \label{fig:random_whole_data}
\end{figure*}

\begin{figure*}
    \hspace*{-0.5cm}\includegraphics[width=1.04\linewidth,clip,trim=0cm 0cm 0cm 0cm]{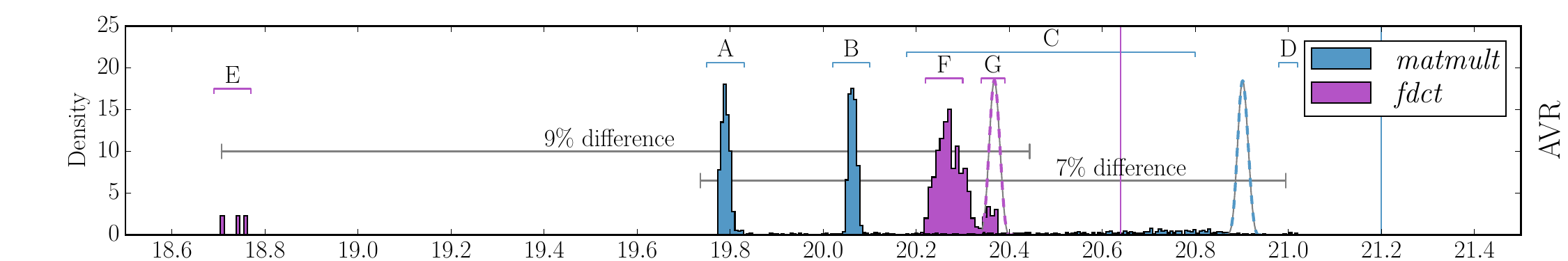}
    \hspace*{-0.5cm}\includegraphics[width=1.04\linewidth,clip,trim=0cm 0.5cm 0cm 0.5cm]{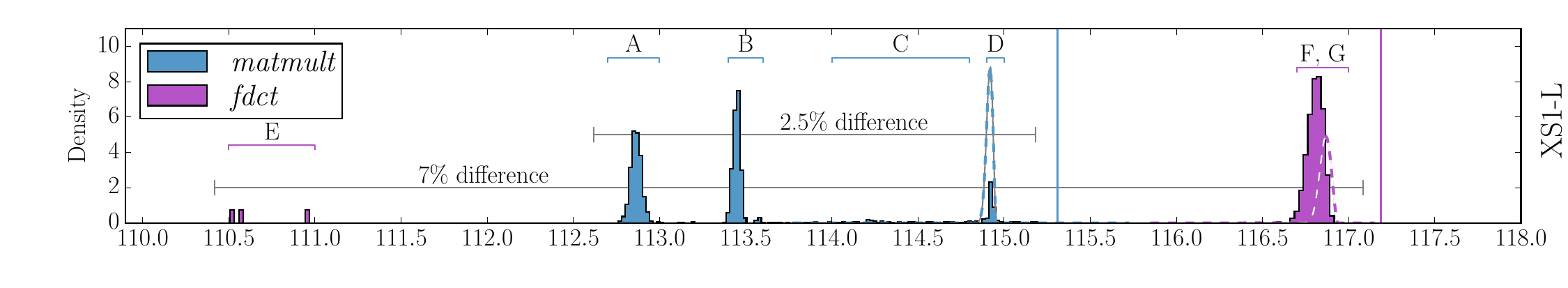}
    \caption{Distribution of benchmark average power (mW), when run with hand-crafted datasets.}
    \label{fig:contrived_whole_data}
\end{figure*}

We wish to validate that our modeling technique can be applied to whole
programs and that it works on more platforms than just the Atmel AVR.  However,
the cost of modeling all the instructions that appear in a full program is
large. Therefore, we measure the energy distribution of full embedded
software programs on different processors to determine whether it can be
characterised
by the Weibull distribution. If so, and similar results are seen on different
processors, then our composition technique should be capable of modeling the
behaviour of a full program.

Additionally, we seek to verify that the probabilistic upper bound of a program
distribution is sound, by using hand-crafted data to try and exercise a range
of data switching behaviours in the processor, and existing GA techniques to
search for inputs with high energy consumption (in the manner of
W\"agemann~\cite{Wagemann2015} and Liqat~\cite{fopara15}).

To demonstrate that this technique (i.e. energy distribution) is not specific
to the Atmel AVR, we test benchmarks on the XMOS \xsonel, a deeply embedded
cacheless processor. Both the AVR and \xsonel are appropriate for IoT
applications. The AVR has an 8-bit data-path, whereas \xsonel is 32 bits. The
\xsonel is a single-core operating at 400~MHz with a four-stage hardware
multi-threaded pipeline.  Using single threaded benchmarks, the pipeline is
only 25\,\% utilised.  However, the effects of data on these benchmarks is
still measurable.
\vspace{3ex}
\subsection{Benchmarks}

For benchmarks, we select programs that have no
data dependent branches, therefore changes in energy are purely due to
different data progressing through the computational path in the processor.
Programs that have data-dependent branches may execute different sequences of
instructions, and may have different execution times, thus we exclude these to
focus purely on changes in energy due to data values, not program flow.
This matter is discussed further in \Cref{s:discuss}.

The benchmarks used for this test are \benchmark{fdct} and
\benchmark{matmult-int}, taken from BEEBS, an embedded benchmark
suite~\cite{Pallister2013b}. These tests are purely integer, because the target
processors in this work have no hardware floating-point support.

\benchmark{fdct} has a state space of 1024 bits (one 8x8 block of 16 bit pixel
data), while \benchmark{matmult-int} performs a 20x20 matrix multiplication of
8 bit integers. Both cannot be explored exhaustively, and thus would benefit
from statistical characterisation.
Neither of the chosen benchmarks have data dependent branches, thus their execution time is identical even with different input data.

\subsection{Random data}

\Cref{fig:random_whole_data} shows the average power when the \benchmark{fdct} and \benchmark{matmult-int} benchmarks use random data. The dashed line shows the Weibull distribution fitted to these data. Overall, the distributions are narrow, indicating a low variation caused by the data. The variations for both benchmarks on AVR are similar, however, each has a different mean, since different instructions are executed, each with a different average power.

Fitting these parameters to the Weibull distribution for each of the
benchmarks results in an estimation of the maximum achievable average power for
each benchmark, as in \Cref{s:modelling}.
For example the probabilistic maximums for AVR are
\SI{20.64}{\milli\watt} for \benchmark{fdct} and \SI{21.20}{\milli\watt} for
\benchmark{matmult}. These bounds are shown by the solid vertical lines in
\Cref{fig:random_whole_data}.

\subsection{Genetic algorithm}

Related work shows that genetic algorithms (GAs) are an effective technique to
find the maximum power dissipation for circuits and software. Our paper
instantiates a GA that attempts to find a dataset which increases the energy or
power for the entire program.

The results are included in \Cref{fig:random_whole_data} as dotted vertical
lines. These data points are slightly higher and lower than the points found by
the random data --- the guidance provided by the GA allows both higher and
lower solutions to be found quickly. Since the parameters to the Weibull
distribution were found for each distribution, the probability of finding a
more extreme solution can be calculated. For our test cases, the probabilities
are less than $10^{-9}$, provided the assumption of the distribution being a
good fit holds. However, the size of the data input space is so large that
there are many possible states which may trigger a larger energy consumption.

\subsection{Hand-crafted data}

Due to the extremely large number of input states, certain configurations of
input are never considered by random search or the GA. This includes data such
as every bit set and every bit cleared, which could be important and trigger an
unusually high or low energy consumption. We hand crafted certain inputs to
exercise these edge cases, to evaluate whether the GA or probabilistic bounds
could be exceeded.  The types of hand-crafted data fed into each benchmark are:

\begin{description}

    \item[All bits zero.] All of the bits in the data values are set to zero.
    \item[All bits one.] All of the bits in the data values are set to one.
    \item[Strided ones.] The data element is set to one at various strides, such as every 2, 4, 8 or 16 bytes.
    \item[Strided rand.] As above, with stride contents randomised.
    \item[Patterns.] Patterns known to cause high energy consumption. E.g. \texttt{0xa..a} and \texttt{0x5..5} in multiplication~\cite{Kerrison2015}.
    \item[Sparse.] Only one element set to one in various positions.
    \item[Restricted bit-width.] Setting random $n$-bit values in a $m$-bit offset region, as shown below:

    \begin{center}
    \begin{tikzpicture}[scale=0.35]
        \foreach \pnt in {0,1,2,3,4,5,6,7,8,9,10,11,12,13,14,15}
        {
            \begin{scope}[xshift=\pnt cm]
            \draw (0,0) -- (0,1) -- (1,1) -- (1,0) -- cycle;
            \end{scope}
        }

        \foreach \pnt in {4,5,6,7,8,9}
        {
            \begin{scope}[xshift=\pnt cm]
            \draw[fill=gray] (0,0) -- (0,1) -- (1,1) -- (1,0) -- cycle;
            \end{scope}
        }

        \draw[latex-latex] (16,-0.5) -- (10, -0.5) node [below, midway] {$n = 6$};
        \draw[latex-latex] (10,-0.5) -- (4, -0.5) node [below, midway] {$m = 6$};
    \end{tikzpicture}
    \end{center}
    \item[All elements the same.] Every element in the data is set to the same value. A range of values are tested.

\end{description}

\Cref{fig:contrived_whole_data} shows the average power when all of these hand-crafted sets of data are measured on each benchmark. There are many different components of these graphs --- each caused by a different part of the hand-crafted data. They are discussed below, starting with \benchmark{matmult}.

\begin{description}

    \item[A] This mode is around the lowest average power achievable for the
      \benchmark{matmult} benchmark, all zero data or sparse data.

    \item[B] The distribution consists of sparse elements with few bits set to one. This causes low average power since at most single bits set to one are multiplied, with repeated zeros in-between.

    \item[C] There are a spread of points at this location, formed from tests
      with more dense data.

    \item[D] The highest consumption observed in the non-sparse tests is 21.04~mW for
      AVR, and is caused by data which has the same value in all the elements.
      The values of the elements for the top results are 247, 253, 181, 221 and
      245 --- close to having all bits set. These are the only tests which
      significantly exceed the distribution obtained from random data. For the
      \xsonel, a larger proportion of tests dissipate a higher power, visible in
      the form of a third peak.
\end{description}

\noindent
There are three modes of interest for \benchmark{fdct}:

\begin{description}

    \item[E] Three data points which are
      far lower than any other. These are all zero data and two instances
      of strided data, when the first in every 32 elements is one and all
      elements are zero. This is sparse data, however any of the other sparse
      data still triggers much higher power. This characteristic is observed on
      both architectures.

    \item[F] The majority of tests occur in this bracket, below the expectation
      given by random data. The AVR is an 8-bit processor, so 16-bit arithmetic
      can require two instructions, for upper and lower bits. Many of the
      hand-crafted data sets use zero or close to zero value data, resulting in
      the upper operation having lower power.  This is not the case with the
      32-bit \xsonel, which produces a single peak.

    \item[G] These tend to be triggered by higher-order bits set.  With high
      valued data, the second (upper) part of the arithmetic operation has
      non-zero value, corresponding to a higher average power on AVR.

\end{description}

\subsection{Analysis}

Overall, there is a trend towards higher average power as data becomes more
random or dense. The distribution predicted by random data is a good estimation
of the upper bound. Several tests exceed the limits found with
GAs, but all are bounded by the probabilistic highest value.

Comparing the characteristics observed on both processors, the distributions
take similar forms for both \benchmark{matmult} and \benchmark{fdct}. The
\xsonel dissipates more power, but is a more complex device with a higher
operating frequency. However, the separation between the distributions A and B
in \benchmark{matmult} are within the same order of magnitude for both devices.
Similarly, the widths of the features denoted F in \benchmark{fdct} differs by
a comparable amount. Overall, for AVR and \xsonel respectively,
\benchmark{matmult} is shown to have a power variation of 6.5\,\% and 2.2\,\%,
with \benchmark{fdct} showing 9.1\,\% and 6.0\,\%. These are within the error
margins of many energy models, and are in fact likely to be a contributing
factor to these errors. These variations, along with environmental factors that
may influence device energy consumption, must be considered in order to
establish upper energy bounds with adequate safety.

In summary, it appears that for both platforms tested, the energy
distribution of a program can be characterised as a statistical distribution,
and further that the probabilistic upper bound exceeds all tests tried. With
full instruction models, our technique should be able to statically determine
this distribution.

\section{Discussion}
\label{s:discuss}

Our technique allows for characterising and bounding the amount of energy
consumption caused by variations in data operands in a program. A complete
analysis requires a model for each instruction pairing for each platform, which, while
costly, is a worthwhile trade-off for accurate offline energy estimation. In comparison
with the technique of Steinke et al. \cite{Steinke2001}, rather than
characterising each instruction by its input and the current processor state,
we instead produce a single probability distribution representing the likely
costs of transitioning from one instruction to another, enabling composition.

However, the technique does not fully account for all energy consumption in a
processor. A large cost in many processors is static leakage, the energy lost
for every moment that the processor is active. Existing WCEC
techniques~\cite{Jayaseelan2006,Wagemann2015,taco:GeorgiouKCE2017} combine both
(an approximation of) circuit switching costs with standard techniques to
estimate worst case code paths, such as IPET~\cite{Li1995a}. To fully account
for processor costs, our probabilistic technique would need to be combined in a
similar fashion to find the greatest probabilistic upper bound on energy on all
paths through a program. We have not yet explored this as we focus only on data
dependent costs here.

Our technique also benefits from the predictable nature of processors suited
for the IoT domain, i.e. those without caches, branch prediction or speculative
execution, all of which have energy costs that heavily depend on program state.
Were the technique to be applied to processors with such features, then
probabilistic models would need developing for those features too, such as the
work of Puranik et al.~\cite{probcaches}, although they focus on the mean
execution time rather than either energy or upper bounds. The tightness of
bounds achievable for such features remains an open question.

Finally our technique does not address external factors in energy consumption,
such as system-level energy or variability due to environmental conditions.
These full-system factors fall well outside the bounds of embedded software
analysis.

\section{Conclusion and future work}
\label{s:conclusion}

This paper has analysed how data values within processors affect energy
consumption. Basic blocks can be modeled by composing instruction models, and
programs with complex control flow can be modeled by composing basic blocks.
To create a composable analysis, the transition between instruction
pairs was modeled as a Weibull distribution. Distributions for instruction
pairs can then be convolved to give a probability distribution of energy for an
instruction sequence.

Several instruction sequences were tested, comparing the predictions to the
actual measured distributions to validate our model. The prediction is tight,
but overestimates the energy consumption in all cases, providing an
estimate of the likely worst case energy consumption. The prediction assumes that all
of the instructions are independent of each other, which is not
generally true.  Repeating the measurements with dependent variables in an
instruction sequence shows that added correlation between the values decreases
the total energy consumption. The prediction (ignoring dependencies) still provides an upper bound as expected in this case, but it is not as tight.

The correlation between data values input and output from instructions can lead
to unusual energy behaviour. An instruction sequence was shown to
produce bimodal energy behaviour across a range of random data, caused by
repeatedly biasing data values towards zero. In such a case of strong
correlation between data values, our model will over-predict.

Initial analysis of full programs suggests that using random
data to create a Weibull distribution allows a probabilistic worst case for
that program to be estimated. This worst case was higher than
could be found using a GA, random or hand-crafted data, giving us confidence
that our estimates are safe.

More generally, this work has shown that worst case energy analysis requires more than
simply a model generated from profiling with random input data. The
distribution of profiling results must be analysed to determine a likely maximum.
Physical system properties such as energy consumption are inherently noisy, and
cannot be as tightly or reliably bound as execution time. However, we
demonstrate empirically that taking a high-order percentile of a Weibull
distribution, fit to random input data, provides a basis for pragmatic WCEC
modeling. We also show that the distributions of energy for instruction
transitions, rather than individual instructions, are necessary when creating a composable model.

A next step towards more accurate static techniques for WCEC estimation
would be to model the entire instruction set using the energy modelling technique presented in this paper, and to combine such a cost model with a worst-case
analysis technique such as IPET, creating a technique that considers both
length and data effects on program energy probabilistically. Following existing
WCEC techniques, this should yield a more accurate upper bound through
consideration of additional energy contributors.

A further observation is that programs have differing degrees of data
dependency --- some instructions in the program are purely control, and do not
operate on the data input to the program. Static analysis could find only the
instructions that are in the data path of the program, and an estimate of the
total variability due to data could be constructed from these instructions and
their transition distributions.

\subsection*{Acknowledgements}

This research has received funding from the European Union 7th Framework
Programme (FP7/2007-2013) under grant agreement no 318337, ENTRA -
Whole-Systems Energy Transparency; grant agreement no 611004,
ICT-Energy; and the ARTEMIS Joint Undertaking under grant agreement 621429,
EMC2. This study was also partly sponsored by EPSRC's Doctoral Training
Account EP / K502996 / 1 (to the first author).

\balance

\bibliographystyle{plain}
\bibliography{bib} %entra}

\end{document}